\documentclass[twocolumn,showpacs,floats,floatfix,superscriptaddress,aps,pra]{revtex4}
\usepackage{amsfonts}
\usepackage{amssymb}
\usepackage{amsmath}
\usepackage{graphicx}
\usepackage{bm}
\usepackage{color}
\usepackage{epsfig}
\usepackage{calc}
\usepackage{ifthen}

\begin{document}

\title{Broadband terahertz half-wave plate with multi-layers metamaterial via quantum engineering}
\date{\today }

\begin{abstract}
In this paper, we employ the novel design of the metamaterial half-wave plate by using the multiple layers of the metamaterials with some specific rotation angles. The rotation angles are given by composite pulse control which is the well-known quantum control technique. A big advantage of this method can analytically calculate the rotation angles and can be easily extended to the multiple layers. The more layers of the metamaterials can present more bandwidth of half-wave plate. In this paper, we present 1, 3, 5, 7 configurations of layers of a metamaterial half-wave plate and we demonstrate that our device increasingly enhance the bandwidth of performance.
\end{abstract}

\author{Wei Huang}
\affiliation{Guangxi Key Laboratory of Optoelectronic Information Processing, Guilin University of Electronic Technology, Guilin 541004, China}

\author{Xiaoyuan Hao}
\affiliation{Guangxi Key Laboratory of Optoelectronic Information Processing, Guilin University of Electronic Technology, Guilin 541004, China}

\author{Yu Cheng}
\affiliation{Guangxi Key Laboratory of Optoelectronic Information Processing, Guilin University of Electronic Technology, Guilin 541004, China}

\author{Shan Yin}
\email{syin@guet.edu.cn}
\affiliation{Guangxi Key Laboratory of Optoelectronic Information Processing, Guilin University of Electronic Technology, Guilin 541004, China}

\author{Jiaguang Han}
\affiliation{ Center for Terahertz Waves and College of Precision Instrument and Optoelectronics Engineering, Tianjin University, Tianjin 3000072, China}

\author{Wentao Zhang}
\email{zhangwentao@guet.edu.cn}
\affiliation{Guangxi Key Laboratory of Optoelectronic Information Processing, Guilin University of Electronic Technology, Guilin 541004, China}
\maketitle



\section{Introduction}
The half-wave plate (HWP) is the essential linear optical device and it widely used in the basically all linear optical systems \cite{Hecht2002, Born2005, Goldstein2003}. The normal (or bulk) HWP constitutes of birefringence crystal, which rotates the polarization vector of an incident light at 2$\theta$, where $\theta$ denotes the angle between the incident light polarization direction and the fast axis of the waveplate \cite{Hecht2002, Born2005, Goldstein2003}. Due to the large development of the metamaterial, recently, HWP can be achieved by metamaterial \cite{Zi2018, Ding2015, Ma2018, Liu2017, Xia2017, Dong2019, Liu2019} which has specific structure of unit cell with much smaller than the wavelength of input light or electromagnetic wave. Comparing with bulk HWP, metamaterial HWP has the advantage of ultra-thin and compact, and also has the good performance at the terahertz (THz) wave. 

The normal all-dielectric THz metamaterial HWP only works at single frequency or narrow bandwidth \cite{Zi2018, Liu2017}, leading to non-universal applications.
Thus, improving the bandwidth of THz HWP largely increases the universal applications of THz technologies, such as the communication \cite{Nagatsuma2016} and spectroscopy \cite{Yang2016}. 
Therefore, enhancing the bandwidth of performance of THz HWP is a very interesting and significant topic. 
There are few designs of broadband THz metamaterial HWP consisting of some complex and special-design structure of the unit cell \cite{Liu2017, Xia2017, Dong2019}. 
However, those devices are very hard to design the complex structure of unit cell and only can perform at the specific bandwidth. By changing the specific bandwidth should re-design the complex structure of unit cell. 
Most recently, a remarkable paper to design the broadband THz HWP by using the multiple layers of metamaterial and the parameters of each layer is given by the inverse design \cite{Liu2019}. The disadvantages of this method are that the corresponding parameters of each layer requires complex numerical calculations and it is very hard to find the pattern to analytically obtain the parameters of each layer.
To overcome these issue, we firstly propose the broadband THz HWP via composite pulse control technique.

\begin{figure*} [htbp]
\centering
\includegraphics[width=0.7\textwidth]{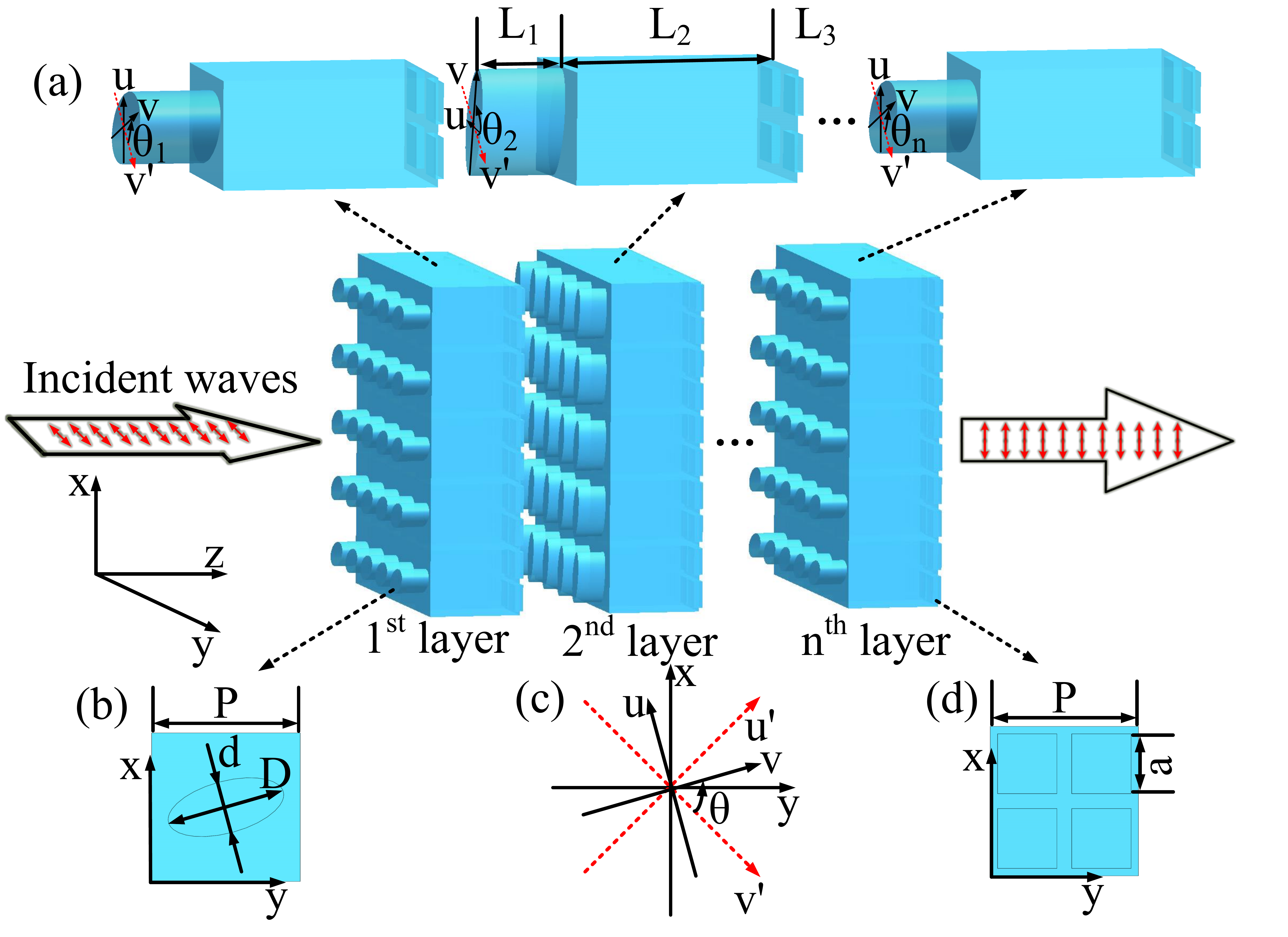}
\caption{The schematic figure of our design. (a) Multiple layers of our device with the rotating unit cells. (b) The cross-section of the elliptical part of unit cell. (c) The single layer of fast and slow axis (red dashed) and corresponding rotation angle $\theta$. (d) The cross-section of the anti-reflection part.} 
\label{fig1}
\end{figure*}

Composite pulse control (a well-known coherent quantum control technique) is originally developed for use in nuclear magnetic resonance (NMR) to suppress the pulses' errors, which replaces one pulse with multiple pulses to increase the robustness of nuclear spin system \cite{Levitt1986, Tycko1985, Cummins2003}. Recently, there are some remarkable papers employing composite pulse control to design broadband and ultra-broadband HWP which replaces the single HWP with the multiple HWPs in the optical light range \cite{Dimova2016, Dimova2015, Dimova20152, Stoyanova2019, Mahmoud2020}. Most recently, some papers have already proposed quantum control to enhance the THz device \cite{Huang2018, Huang2019, Huang2020}. Therefore, we propose novel broadband THz HWP by combining composite pulse control via multiple layer THz HWP (see Fig.~\ref{fig1}). 

In this paper, each layer of HWP consists of three parts of the dielectric materials, such as the elliptical material (the long length $D$ as the fast axis $v$ and the short length $d$ as the slow axis $u$) with length $L_1$ (as shown in Fig. ~\ref{fig1} (a)), the square substrate with the length $L_2$ and four small square dielectric material (length of side $a$) as the anti-reflection layer with the length $L_3$ (see Fig. ~\ref{fig1} (d)), which was proposed by ref. \cite{Zi2018}. In our paper, each layer has the same structure from the ref. \cite{Zi2018}, but elliptical material part of each layer requires the specific anti-clock rotation (see Fig.~\ref{fig1} (c)). The unit cells of multiple layers have shown in Fig. ~\ref{fig1} (a) with the different rotation angles from the $\theta_1$ to the $\theta_n$. In order to measure the performance of our device, the polarization of incident light is y-polarized THz wave and we measure the output wave of 'x-polarized' THz wave to obtain the PCR. 

In this paper, we firstly demonstrate the theory of Jones matrix of metamaterial THz HWP and composite pulse control. Subsequently, we obtain the rotation angles of each layer, given by composite pulse control (see Table~\ref{Table1}) and theoretically predict the performance of our device as shown in Fig.~\ref{fig2}. Finally, we demonstrate the performance of our device in full-wave simulation to further illustrate the performance of our design (see Fig.~\ref{fig3} and \ref{fig4}).

\section{Theory}
In this paper, we employ the signal layer of all-dielectric metamaterial HWP as shown in ref. \cite{Zi2018}, which consists of three different parts, such as the elliptical (with $D =  80 \mu m$, $d = 35 \mu m$ and length $L_1 = 300 \mu m$),  the substrate (the length $L_2 = 1000 \mu m$) and the anti-reflection part (four small square with $a = 45 \mu m$ and the thickness $L_3 = 20 \mu m$), as shown in Fig. ~\ref{fig1}. In the Jones calculus a single layer of metamaterial rotated at an angle $\theta$ with reference to the slow and the fast axes of the plate is described by the Jones matrix,
\begin{equation}
\mathfrak{J}_{\theta }(\varphi )=\mathfrak{R}(-\theta )\left[
\begin{array}{cc}
e^{i\varphi /2} & 0 \\
0 & e^{-i\varphi /2}%
\end{array}%
\right] \mathfrak{R}(\theta ),
\end{equation}%
where
\begin{equation}
\mathfrak{R}(\theta )=\left[
\begin{array}{cc}
\cos \theta & \sin \theta \\
-\sin \theta & \cos \theta%
\end{array}%
\right] ,
\end{equation}%
and the phase retardance is $\varphi$. The Jones matrix represents a half-wave plate, when the phase retardance is $\varphi =\pi $. Therefore, we use this Jones matrix to represent the single layer of metamaterial HWP. 


\begin{table}[t]
\begin{tabular}{|c|l|}
\centering
$N$ & Rotation angles $(\theta _{1}$;\; $\theta _{2}$;\; $\cdots $;\; $%
\theta _{N-1}$;\; $\theta _{N}$) \\ \hline
3 & (60;\;120;\;60) \\ \hline
5 & (51.0;\;79.7;\;147.3;\;79.7;\;51.0) \\ \hline
7 & (68.0;\;16.6;\;98.4;\;119.8;\;98.4;\;16.6;\;68.0) \\ \hline
9 & (99.4;\;25.1;\;64.7;\;141.0;\;93.8;\;141.0;\;64.7;\;25.1;\;99.4) \\
\hline
\end{tabular}%
\caption{Rotation angles $\protect\theta _{k}$ (in degrees) for modular
broadband half-wave plates with different numbers $N$ of constituent
half-wave plates. }
\label{Table1}
\end{table}

\begin{figure} [htbp]
\centering
\includegraphics[width=0.5\textwidth]{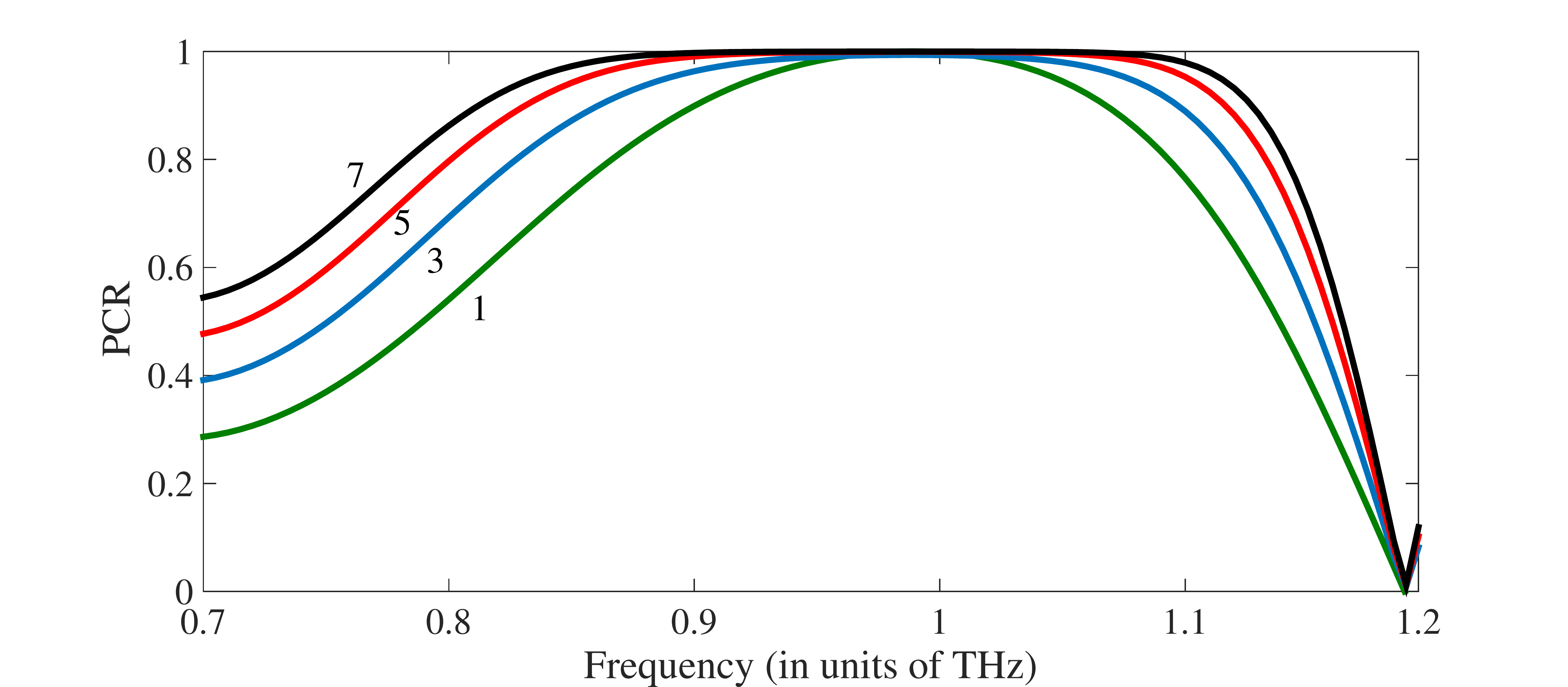}
\caption{The theoretical calculations of $PCR$ against frequency of our broadband THz HWP based on multiple layer, where the green, blue, red and black lines are correspond to single, three, five and seven layer(s) of metamaterial THz HWP. } 
\label{fig2}
\end{figure} 

Subsequently, transfer the horizontal-vertical polarization (HV) to Left-right circular polarization (LR) basis, by employing $\mathbf{J}_{\theta }(\varphi )=\mathbf{W}^{-1}\mathfrak{J}_{\theta }(\varphi )\mathbf{W}$, where $\mathbf{W}$ is given by
\begin{equation}
\mathbf{W}=\tfrac{1}{\sqrt{2}}\left[
\begin{array}{cc}
1 & 1 \\
-i & i%
\end{array}%
\right] .
\end{equation}%
Thus, the Jones matrix for single layer metamaterial HWP with a phase retardance $\varphi$ and rotated by an angle $\theta $ is given as (in the LR basis),
\begin{equation}
\mathbf{J}_{\theta }(\varphi )=\left[
\begin{array}{cc}
\cos \left( \varphi /2\right)  & i\sin \left( \varphi /2\right) e^{2i\theta }
\\
i\sin \left( \varphi /2\right) e^{-2i\theta } & \cos \left( \varphi
/2\right)
\end{array}%
\right] .  \label{retarder}
\end{equation}

Based on composite pulse control \cite{Dimova2016, Dimova2015, Dimova20152, Stoyanova2019, Mahmoud2020, Ivanov2012, Ardavan2007}, we replace multiple layers of metamaterial HWP to single layer metamaterial HWP with the corresponding rotation angle $\theta _{k}=\theta _{N+1-k}$, $(k = 1, 2, ..., n)$. Therefore the Jones matrix of multiple layers of metamaterial HWP is given by, 

\begin{equation}
\mathbf{J}^{\left( N\right) }=\mathbf{J}_{\theta _{N}}\left( \varphi \right)
\mathbf{J}_{\theta _{N-1}}\left( \varphi \right) \cdots \mathbf{J}_{\theta
_{1}}\left( \varphi \right) .  \label{Jn}
\end{equation}%

We aim to implement an ideal half-wave plate propagator with Jones matrix $\mathbf{J}_{0}$ on the LR basis (up to a global phase factor),
\begin{equation}
\mathbf{J}_{0}=\left[
\begin{array}{cc}
0 & i \\
i & 0%
\end{array}%
\right] ,
\end{equation}%
through the modular half-wave plate propagator with Jones matrix $\mathbf{J}^{\left( N\right) }$ from Eq. 
\eqref{Jn}. That is, we set $\mathbf{J}^{(N)}\equiv \mathbf{J}_{0}$ at $%
\varphi =\pi $. We nullify as many lowest order derivatives of $\mathbf{J}^{(N)}$ vs the $\varphi $ at $\varphi =\pi $ as possible, obtained by 
\begin{eqnarray}
\left[ \partial _{\varphi }^{k}\mathbf{J}_{11}^{\left( N\right) }\right]
_{\varphi =\pi } =0, \quad \left[ \partial _{\varphi }^{k}\mathbf{J}_{12}^{\left( N\right) }\right]
_{\varphi =\pi } =0, 
\label{nullify}
\end{eqnarray}
where $(k = 1, 2, ..., n)$. The anagram symmetry assumption for the angles $\theta _{k}$ ($\theta
_{k}=\theta _{N+1-k}$), ensures that all even-order derivatives of $\mathbf{J}_{11}^{\left( N\right) }$ and all odd-order derivatives of $\mathbf{J}_{12}^{\left( N\right) }$ vanish. One well-known analytical solution is given in Table~\ref{Table1}, which has already demonstrated in ref.~\cite{Dimova2016}.

In order to demonstrate the transfer efficiency, we define the fidelity (polarization conversion rate (PCR) in optical language) of the Jones matrix of the modular half-wave plate, such as  
\begin{equation}
PCR = \frac{1}{2}tr (J_0^{-1} J^{(N)}).
\label{fidelity}
\end{equation} 
Based on above theory, we can theoretically calculate the multiple layers of metamaterial HWP $PCR$ via a single layer of metamaterial HWP, as shown in the Fig.~\ref{fig2}. From the calculations, it is very easy to obtain that the increasing number of layers can produce more bandwidth of the performance.

\begin{figure} [htbp]
\centering
\includegraphics[width=0.43\textwidth]{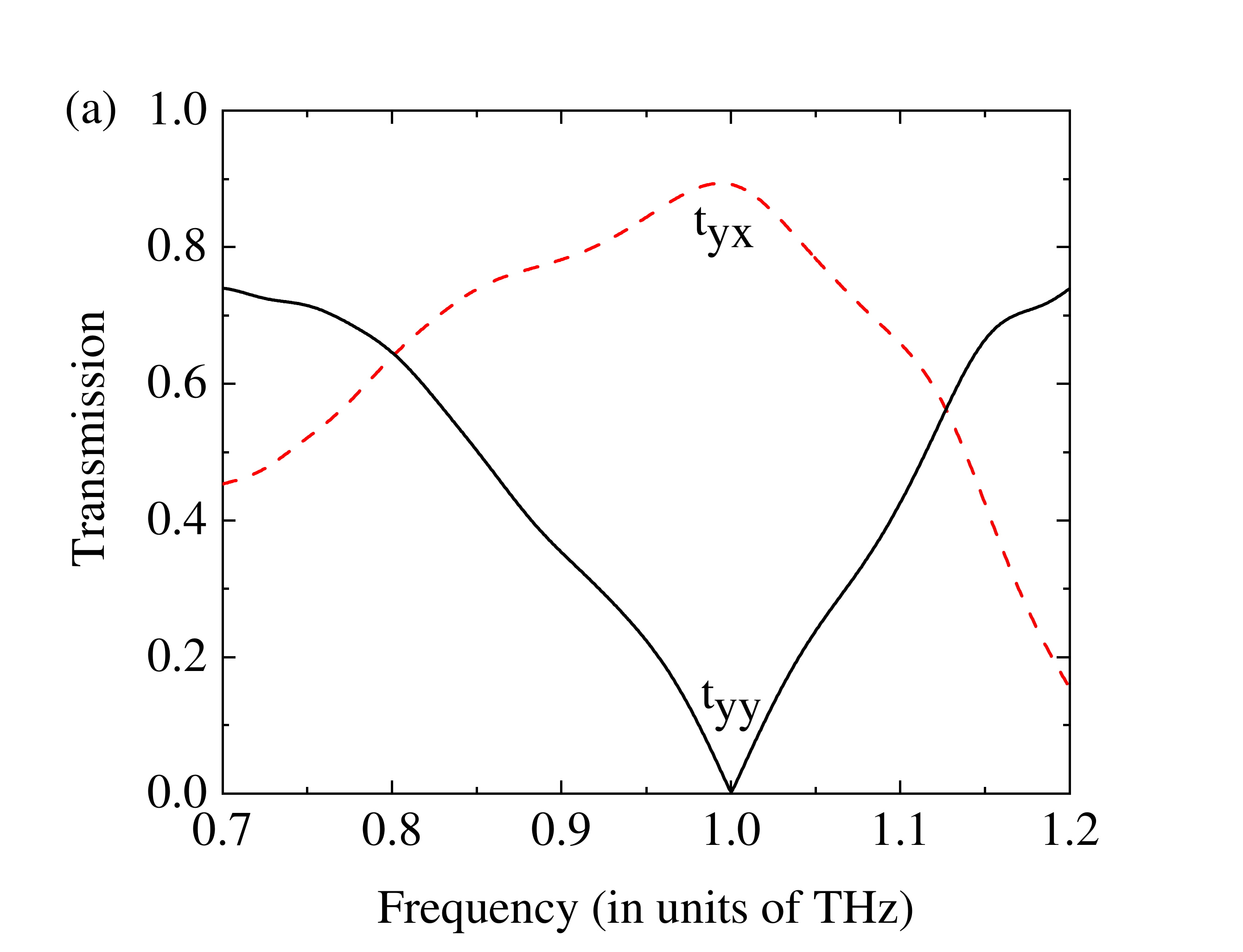}
\includegraphics[width=0.43\textwidth]{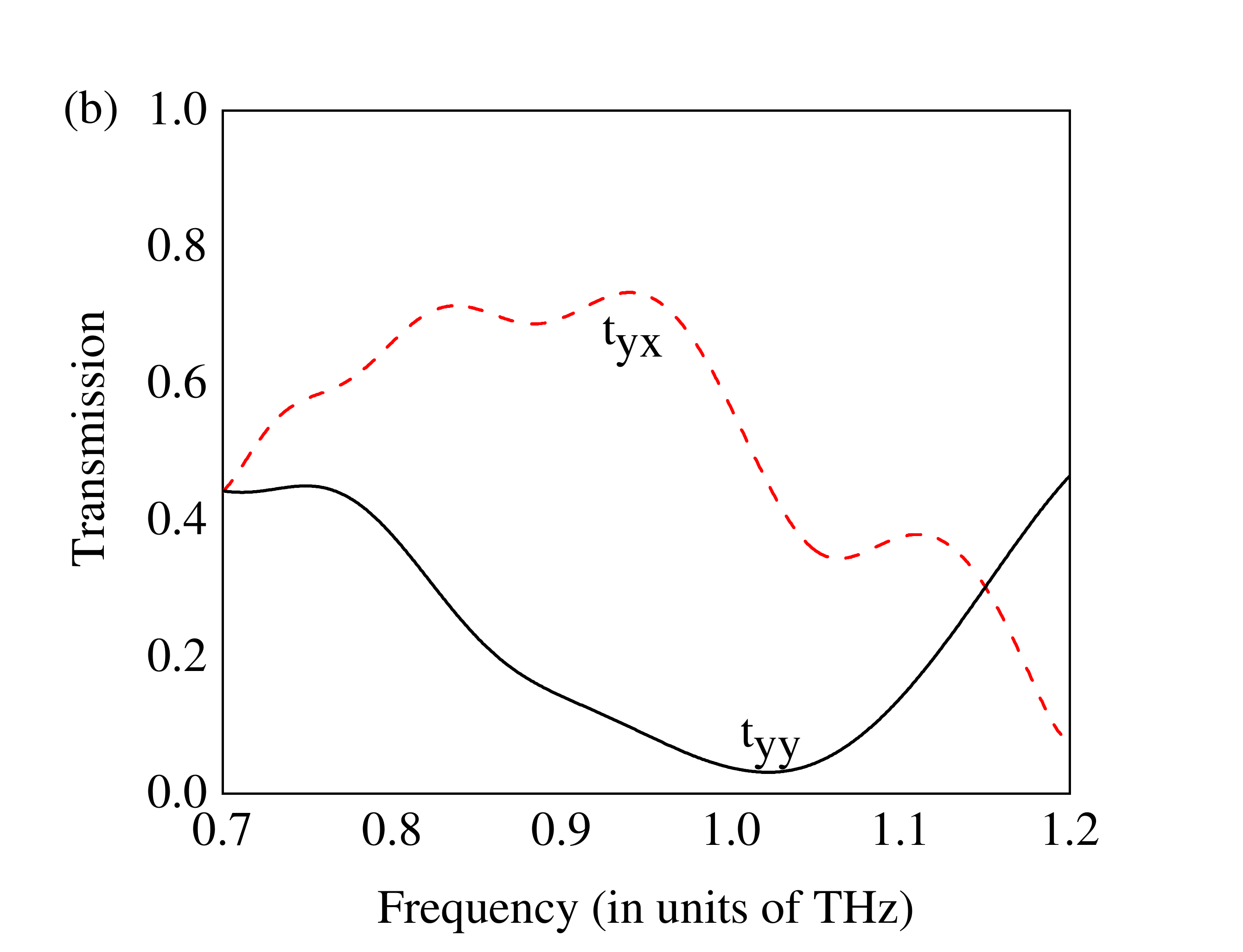}
\includegraphics[width=0.43\textwidth]{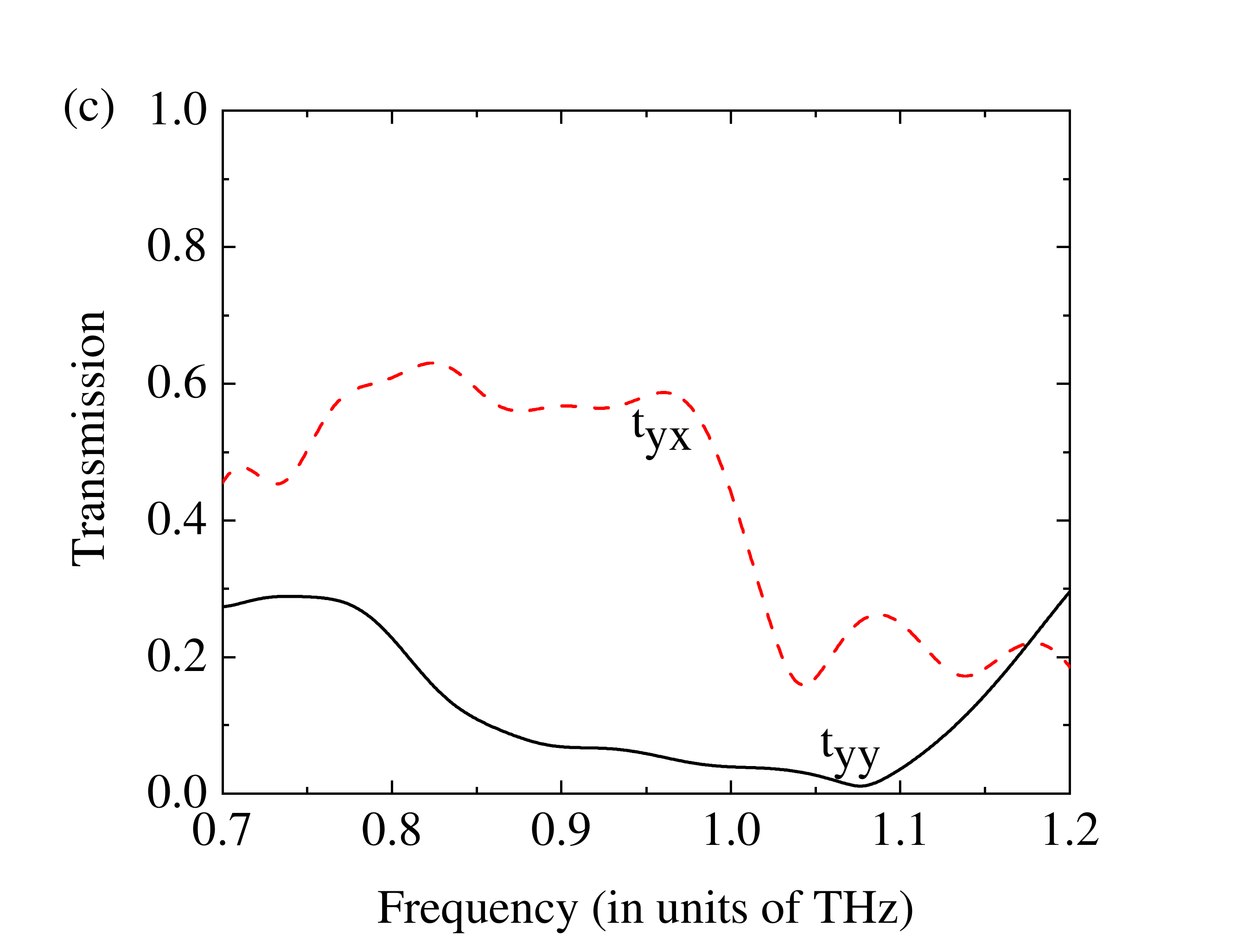}
\includegraphics[width=0.43\textwidth]{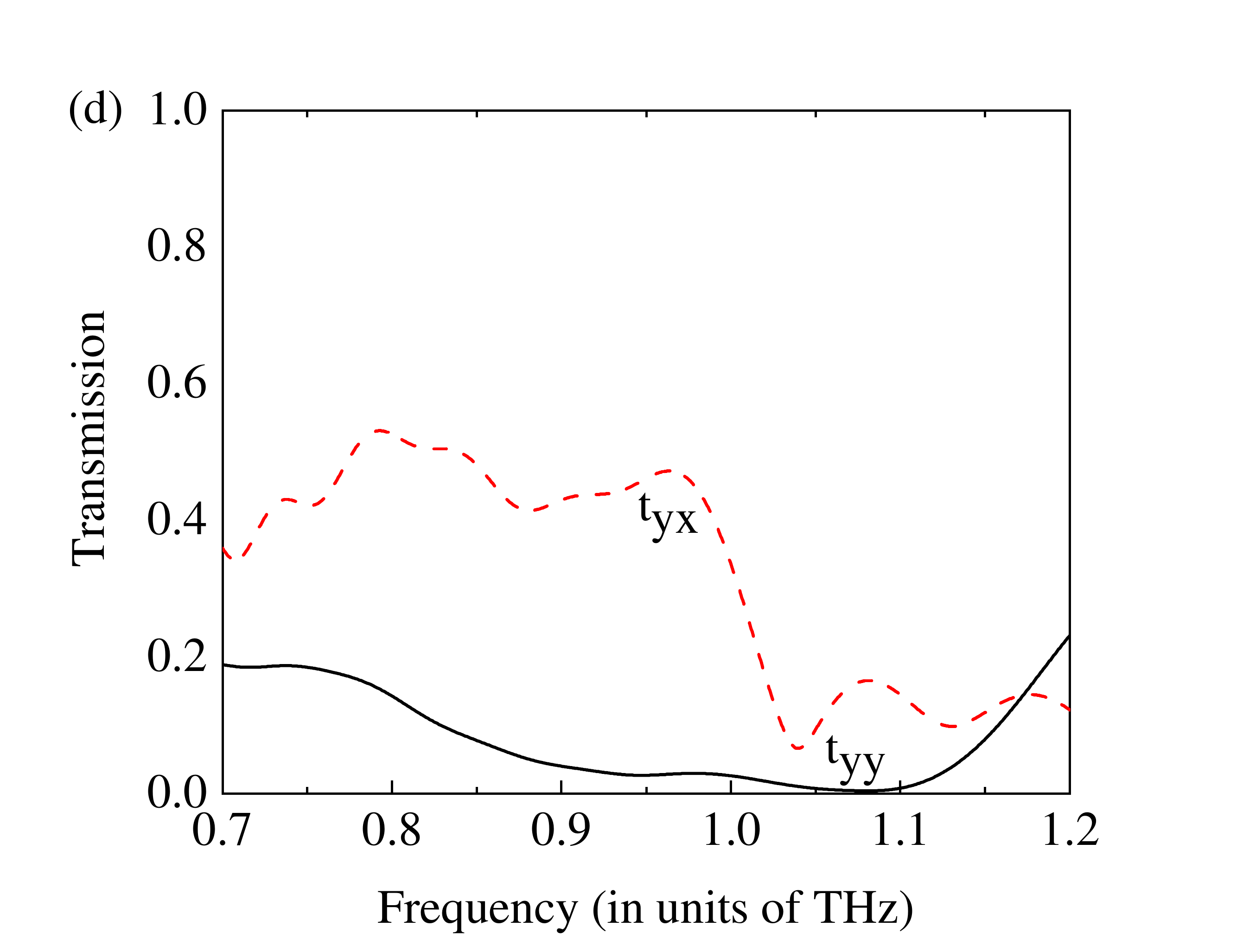}
\caption{The transmission rate $t_{yx}$ (and $t_{yy}$ of 'y'-polarization input and 'x'-polarization (and 'y'-polarization) output, with corresponding to (a) single, (b) three, (c) five and (d) seven layer(s) of metamaterial THz HWP, respectively. } 
\label{fig3}
\end{figure} 

\section{Full-wave simulation}
In order to illustrate the performance of our device, we demonstrate our device with full-wave simulation which is more reliable and realistic. In this section, we demonstrate the output polarization of our device with the single, three, five, seven layer(s) respectively, as shown in Fig.~\ref{fig3}, where $t_{yx}$ (and $t_{yy}$) represents the transmission rate of 'y'-polarization input and 'x'-polarization (and 'y'-polarization) output. The polarization conversion rate ($PCR$) can be defined as, 
\begin{equation}
PCR = \dfrac{t_{yy}^2}{t_{yx}^2+t_{yy}^2}. 
\end{equation}

\begin{figure} [htbp]
\centering
\includegraphics[width=0.45\textwidth]{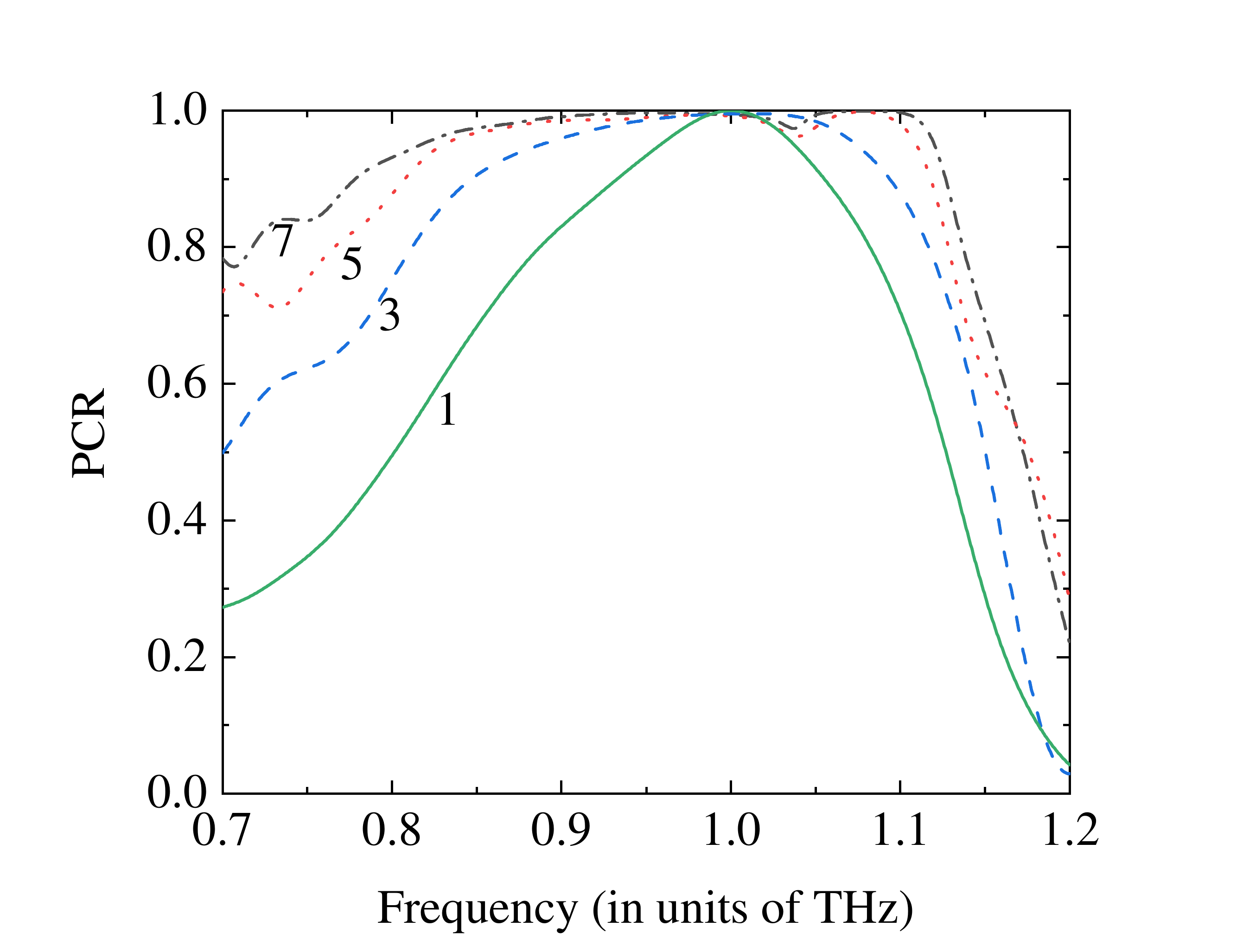}
\caption{The transmission rate $t_{yx}$ (and $t_{yy}$ of 'y'-polarization input and 'x'-polarization (and 'y'-polarization) output, with corresponding to (a) single, (b) three, (c) five and (d) seven layer(s) of metamaterial THz HWP, respectively. } 
\label{fig4}
\end{figure} 

From the results, it is very easy to obtain that our device suppresses the 'y'-polarization output transmission ($t_{yy}$) and transfer to 'x-polarization' output transmission ($t_{yx}$). By increasing the number of metamaterial THz HWP layers, our device can perform more bandwidth of $t_{yx}$. Therefore, we can enhance the bandwidth of the performance by increasing the number of layers via composite pulse control. As we can see in the Fig.~\ref{fig3}, the transmission rates of $t_{yx}$ and $t_{yy}$ are consistent with result in ref. \cite{Zi2018} and $t_{yx}$ has roughly the $90 \%$ transmission rate at the peak. Due to this non-perfect transmission, the multiple layers of metamaterial THz HWP have the lower transmission rate, such as roughly $75 \%$ for three layers, $60 \%$ for five layers and $48 \%$ for seven layers. 

Finally, we demonstrate the $PCR$ for multiple layers of metamaterial THz HWP with single (green solid), three (blue dashed), five (red dotted), seven (black dashed) layer(s) in the full-wave simulations, as shown in Fig.~\ref{fig4}. As we can obtain that the $PCR$ of single layer metamaterial THz HWP is the same results in ref. \cite{Zi2018}, which only can roughly provide the bandwidth ($PCR$ over $90\%$) from  0.933 THz to 1.05 THz. The three layers of our device via composite pulse control can achieve the bandwidth from 0.85 THz to 1.1 THz, which enhances the $114\%$ bandwidth comparing to single layer. Increasing number of our design's layers can provide much more bandwidth, such that the five layers and the seven layers of our device can increase the bandwidth of $164\%$ and $201\%$ respectively. Therefore, our device can increase the bandwidth of metamaterial THz HWP via composite pulse control by analytical calculations of rotation angles. 

\section{Conclusion}
In this paper, we provide a novel design of broadband metamaterial THz HWP by employing multiple layers and the structure of each layer is the same with different rotation angles which are analytically given by composite pulse control (a well-known quantum control technique). Our design enhances the bandwidth ($PCR$ over $90\%$) from $114\%$ (three layers) to $201\%$ (seven layers) comparing to the single metamaterial THz layer.

\section*{Acknowledgements}
This work acknowledges funding from National Key Research and Development Program of China (2019YFB2203901); National Science and Technology Major Project (grant no: 2017ZX02101007-003); National Natural Science Foundation of China (grant no: 61565004; 61965005; 61975038; 62005059). The Science and Technology Program of Guangxi Province (grant no: 2018AD19058). W.H. acknowledges funding from Guangxi oversea 100 talent project; W.Z. acknowledges funding from Guangxi distinguished expert project.


\end{document}